\documentclass[12pt]{article}

\usepackage{newtxtext,newtxmath}

\usepackage{graphicx}
\usepackage{svg}
\usepackage[letterpaper,margin=1in]{geometry}
\usepackage{physics}
\renewenvironment{abstract}
	{\quotation}
	{\endquotation}

\date{}

\makeatletter
\renewcommand{\fnum@figure}{\textbf{Figure \thefigure}}
\renewcommand{\fnum@table}{\textbf{Table \thetable}}
\makeatother

\usepackage{scicite}
\usepackage[margin=-0.65cm]{caption}
\usepackage{url}

\usepackage{todonotes}

\def\scititle{
 Electrically tunable photon correlations from waveguide dipolar polaritons with over a micrometer blockade radius
}
\def\scititle{
 Electrically tunable quantum correlations of dipolar polaritons with micrometer-scale blockade radii
}

\title{\bfseries \boldmath \scititle}

\author{
	Yoad~Ordan$^{1\dagger}$,
	Dror~Liran$^{1\dagger}$,
        Kirk~W.~Baldwin$^{2}$,
        Loren~Pfeiffer$^{2}$,
        Hui~Deng$^{3}$,\and
	Ronen~Rapaport$^{1\ast}$\and
	\small$^{1}$Racah Institute of Physics, The Hebrew University of Jerusalem, Jerusalem 9190401, Israel.\and
        \small$^{2}$Department of Electrical Engineering, Princeton University, Princeton, NJ 08544, USA.\and
	\small$^{3}$University of Michigan, Ann Arbor, MI 48109, USA.\and
	\small$^\ast$Corresponding author. Email: ronenr@phys.huji.ac.il\and
	\small$^\dagger$These authors contributed equally to this work.
}

\begin{document} 

\maketitle

\begin{abstract} \bfseries \boldmath

An extreme yet reconfigurable nonlinear response to a single photon by a photonic system is crucial for realizing a universal two-photon gate, an elementary building block for photonic quantum computing. 
Yet such a response, characterized by the photon blockade effect, has only been achieved in atomic systems or solid states ones that are difficult to scale up. Here we demonstrate electrically tunable partial photon blockade in dipolar waveguide polaritons on a semiconductor chip, measured via photon-correlations.
Remarkably, these "dipolar photons" display a two-orders-of-magnitude 
stronger nonlinearity
compared to unpolarized polaritons, with an extracted dipolar blockade radius up to more than 4~$\mu$m, significantly larger than the optical wavelength, and comparable to that of atomic Rydberg polaritons. Furthermore, we show that the dipolar interaction can be electrically switched and locally configured by simply tuning the gate voltage. Finally we show that with a simple modification of the design, a full photon blockade is expected, setting a new route towards scalable, reconfigurable, chip-integrated quantum photonic circuits with strong two-photon nonlinearities. 
\end{abstract}

\noindent

Photonic quantum computers based on on-chip integrated optical circuits \cite{rudolph2017optimistic} hold great promise due to the inherent speed and protected coherence of photonic qubits and the far superior optical chip scalability compared to competing technologies. A major challenge in universal photonic computing is the lack of sufficient non-linearity to allow for enough photon-photon interactions required for two-photon gates, a necessary element in circuit based computation. Such strong photon nonlinearity is also required for ultrafast electronic control and feedback for tuning and reconfiguring computing elements. A major milestone towards the realization of universal two-photon gates is the observation of two-photon blockade, which is an extreme nonlinear response of a photonic system that only accepts one photon but rejects two (or the opposite effect, an anti-blockade, allowing only two photons but rejecting one). Since bare photons have no direct mutual interaction, such a response can only be achieved through coupling photon to a nonlinear medium and having matter-mediated interactions of the range of the photon wavelength. Establishing such a large interaction length, named the blockade radius, is a challenge.

Photon blockade was first established in single-trapped atom-cavities \cite{birnbaum2005photon} and, in solids, in localized point-like excitations in high-Q optical cavities, such as quantum dots and defect centers in diamond and superconducting junctions \cite{reinhard2012strongly,sipahigil2016integrated, faraon2008coherent,lang2011observation,hoffman2011dispersive}, where the extremely narrow linewidth of the transition in ultrahigh-Q cavities affords blockade with a moderate interaction strength.  However, scaling up of these delicate systems is extremely challenging.  In contrast, photon blockade is more readily achieved with Rydberg atoms with multi-micrometer-sized radii and strong long-range dipolar interactions \cite{dudin2012strongly,peyronel2012quantum}. Such Rydberg-atom based systems are currently strong runners in the race of quantum processing \cite{saffman2010quantum}, but they are limited in physical qubit transport speed, integration complexity, relatively slow gate times, and the challenge of integration into a photonic chip. 
Alternatively, a promising solid-state micrometer-sized system is exciton-polaritons, formed by strong coupling between a photon and a bound electron-hole pair (an exciton- $X$) \cite{deng2010exciton},  which can interact through the exciton component. But unpolarized, neutral exciton-polaritons only interact through a weak contact-like interaction \cite{amo2009superfluidity,ferrier2011interactions}. Even with state-of-the-art devices featuring a linewidth of $<50~\mu$eV, the corresponding  blockade radius is smaller than $0.$~35$~\mu$m, still requiring subwavelength optical confinement in ultrahigh-Q cavities for blockade \cite{delteil2019towards,munoz2019emergence}. Furthermore, no simple electronic or optical control over the process is available.

Recently, however, it has been shown that electrically gated exciton polaritons in scalable on-chip optical waveguides \cite{rosenberg2016electrically} have strong effective dipole-dipole interactions that can be electrically controlled in both space and time \cite{rosenberg2018strongly,liran2024electrically,Dror_unpublished,cristofolini2012coupling,suarez2021enhancement,togan2018enhanced,tsintzos2018electrical}. Therefore, these dipolar polaritons may overcome the severe challenges facing unpolarized polaritons in this context. 
Here we demonstrate an on-chip photonic circuit element based on \textit{dipolar} waveguide polaritons in a semiconductor quantum structure, which are electrically polarized using a local electric gate. These dipolar waveguide polaritons display a clear signature of a two-photon blockade at extremely low densities, with an electrically tunable blockade radius that can become significantly larger than the optical wavelength, and  comparable to that of Rydberg atom polaritons. These results confirm the very large interaction enhancements of electrically polarized exciton-polaritons \cite{rosenberg2018strongly,liran2024electrically}. With simple modifications to the device, we predict that a full blockade can be readily reached, paving the way for \textit{electrically reconfigurable} quantum-optical circuitry with a strong two-photon nonlinearity, in an on-chip planar waveguide geometry \cite{nigro2022integrated,scala2024deterministic}.

\begin{figure}[!thb]
    \centering
        \includegraphics[width=0.7\linewidth]{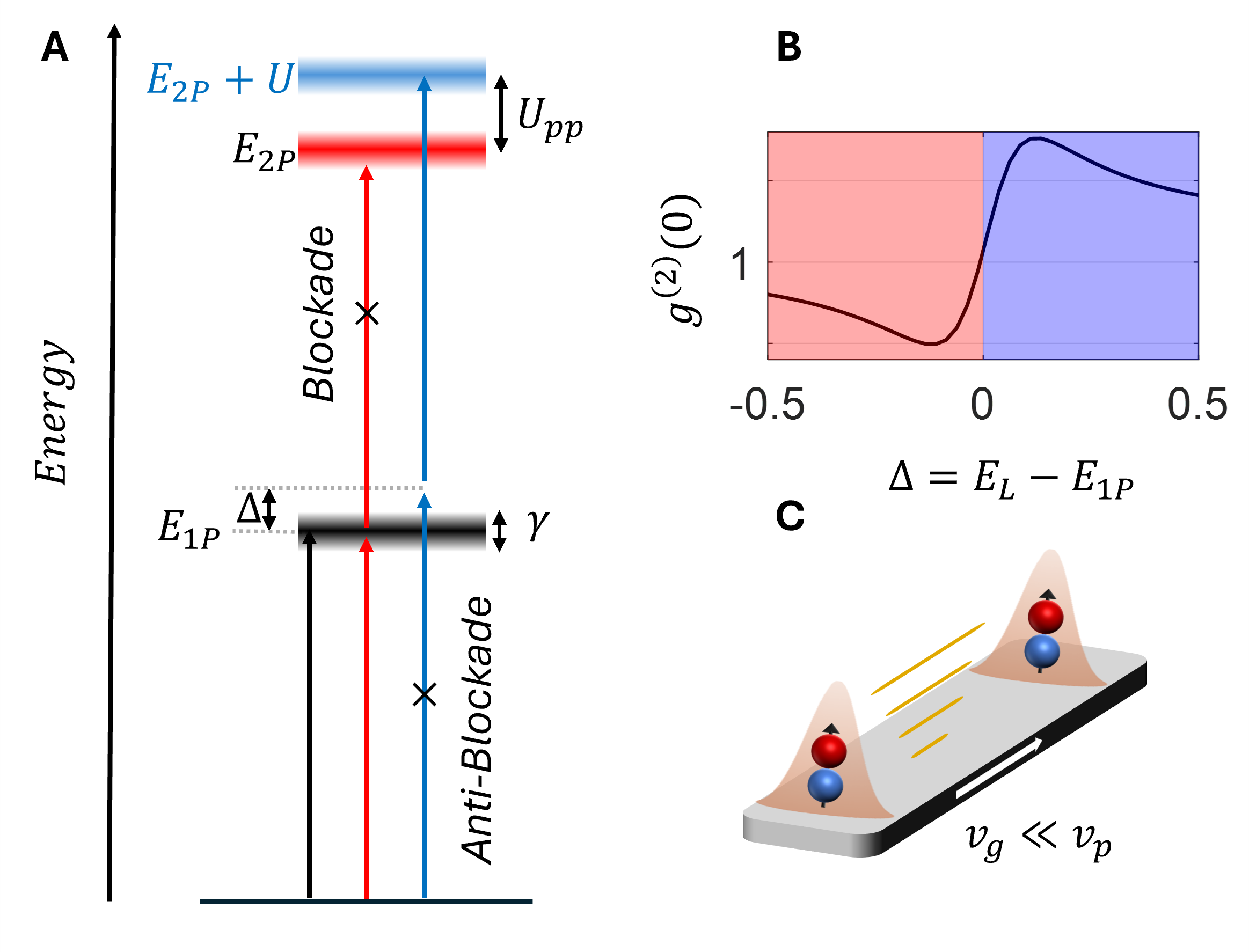}
    \caption{\textbf{Polariton-Polariton interactions and polariton blockade and anti-blockade.}
    \textbf{ (A)} Energy level diagrams for one and two polariton states with a polariton-polariton interaction larger than the natural linewidth of the polariton state, i.e., $U_{pp}>\gamma$, resulting in a selective one polariton (1P) transition (photon Blockade) for red-detuned photons ($\Delta\equiv E_L-E_{1P} <0$), and a selective two polaritons (2P) transition (anti-blockade) for blue-detuned photons ($\Delta >0$). $E_L$ is the energy of the incoming photons.
    \textbf{ (B)} The  expected dependence of the zero-delay second order correlation function of the polaritons ($g^{(2)}(\tau=0$))  as a function of $\Delta$. The red and blue regions mark the blockade regime, manifested in a photon anti-bunching ($g^{(2)}(0)<1$),  and anti-blockade regime manifested in a photon bunching ($g^{(2)}(0)>1$).%
    \textbf{ (C)} schematics of the repulsive interaction of two propagating dipolar exciton-polaritons in an optical waveguide, having a reduced group velocity $v_g$ compared to the bare photon velocity $v_{p}$, due to a large excitonic component $|\chi_x|^2$.
    } 
 
    \label{fig:conceptual}
\end{figure}

Due to their matter component, polaritons can have a mutual interaction $U_{PP}$. If $U_{PP}$ is larger than $\gamma$, the natural linewidth of each polariton excitation, i.e. $U_{PP}>\gamma$,  then only one polariton state is allowed to be excited by an incoming photon having energy of a single polariton transition (1P). On the contrary, two identical resonant photons are blocked from exciting a 2-polariton state (2P), due to the additional energy shift, $U_{PP}$. This blockade scenario is depicted by the red arrows in Fig. \ref{fig:conceptual}A. The blue arrows show the complementary effect, an anti-blockade, where only two photons are allowed but not one.

 A photon blockade (and anti-blockade) can be measured through the zero-delay second-order correlation function, $g^{(2)}_0\equiv g^{(2)}(\tau=0)$ of the outgoing photons  \cite{verger2006polariton}, for different detunings between the energy of the incoming photons and that of the 1P transition $\Delta=E_L-E_{1P}$: if the coupled system is resonantly excited with an incoming weak coherent light having a Poissonian photon number distribution, then the hallmark of the blockade - anti-blockade effect is manifested by a photon anti-bunching at negative $\Delta$ values ($g^{(2)}_0<1$, indicating a lower probability of two simultaneously outgoing photons) followed by bunching behavior at positive $\Delta$ (with $g^{(2)}_0>1$ indicating a higher probability of two outgoing photons). This is shown schematically in Fig. \ref{fig:conceptual}B. Importantly, for a successful blockade, the effective photon-photon interaction should be extensive in space, so that it prohibits two excitations throughout the confining volume of the photons. This condition can be expressed by requiring that the blockade radius $R_b$, defined as: 
 \begin{equation}
    U_{PP}(R_b)=\gamma,
    \label{eq:blockade}
 \end{equation}
 is larger than the distance between any two excitations.

\begin{figure}[!thb]
    \centering
    \includegraphics[width=0.67
    \linewidth]{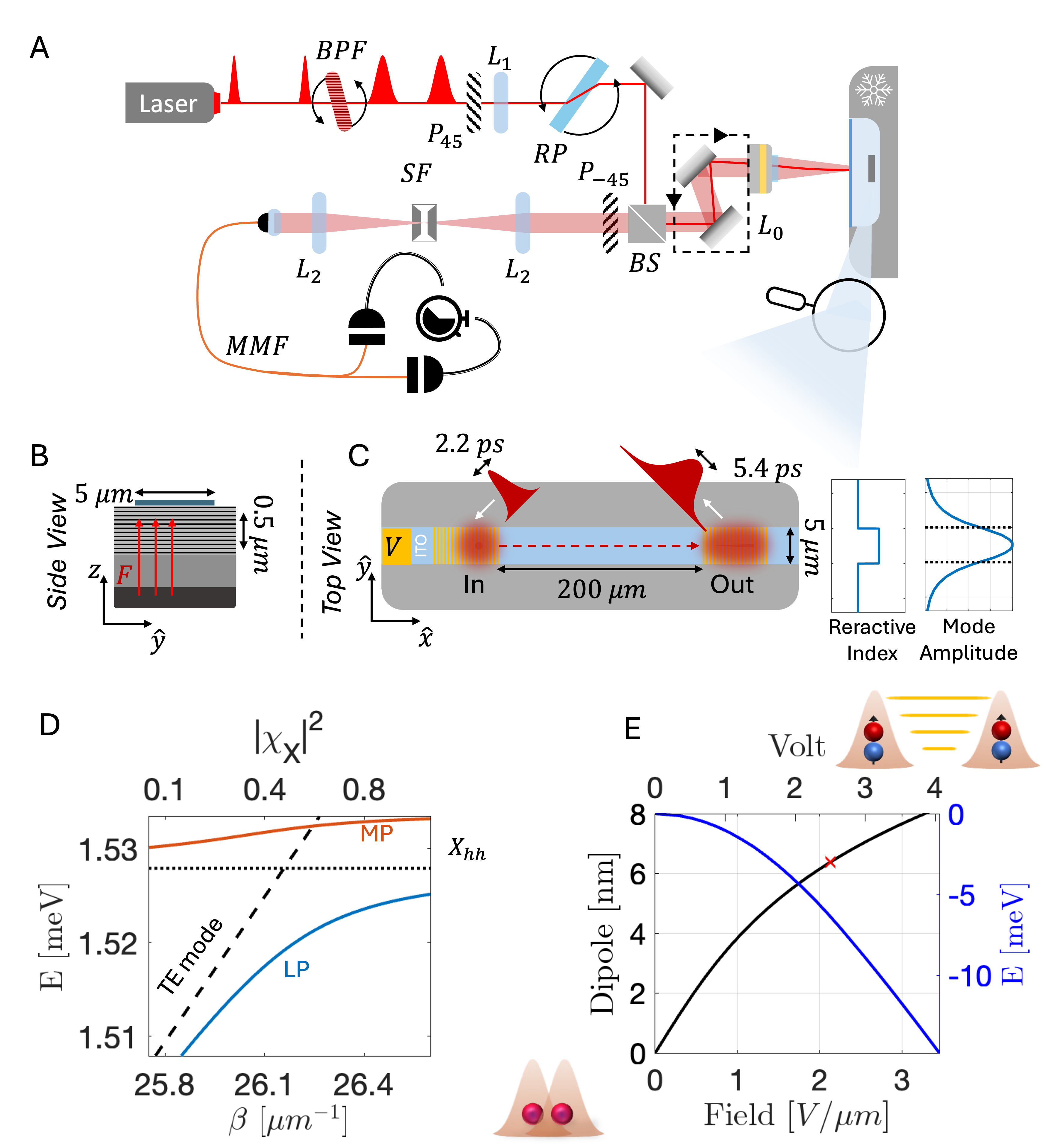}
    \caption{
    \textbf{WG sample and experimental Setup.}
    \textbf{(A)} The setup consists of a  pulsed laser (100fs, 80MHz, less than one photon per pulse on average) focused on the sample (found at $T=5$K).
    The angle of incident is controlled by a Rotating plate (RP). The wavelength and pulse duration is controlled by a rotating band-pass filter (BFP). The cross polarized emission (\(P_{45/-45}\)) from the output grating is selected by a spatial filter (SF) and imaged into a multi-mode fiber (MMF)-based two-photon correlation setup.
    %
    \textbf{(B)} A cross section of the \(5 \mu m\)-wide waveguide (WG) structure. Top blue: ITO, Black horizontal lines: the WG core having 12 GaAs QWs. Bottom black: n$^+$-doped GaAs. Voltage is applied between the ITO strip and the substrate, resulting in electric field lines depicted by red arrows.
    \textbf{(C)} A top view of the biased WG, defined by a 200$\mu$m long conductive and transparent ITO strip.  A laser pulse ($\tau_p=2.2$ps) is focused on the left grating, resonantly injecting polaritons, which couple out from the right grating with a measured temporal length of $\sim5.4 ps$.
    The refractive index profile and the calculated optical mode are also shown. 
    \textbf{(D)} The calculated dispersion of the two lowest exciton-polariton modes (Blue line - lower polariton (LP), Red line - middle polariton (MP), resulting from the strong coupling of the TE mode (dashed line) and the heavy-hole exciton ($X_{HH}$, dotted line), vs. the WG wave-vector $\beta$. The top axis shows the exciton fraction of the LP-mode $|\chi_x(\beta)|^2$ 
    \textbf{(E)} Calculated $X_{HH}$ dipole length (black line), and its energy red-shift (blue line) vs. voltage (top axis) or electric field (bottom axis). The red '$\times$' marks the experimental values.
    }
    \label{fig:Fig2}
\end{figure}

To achieve photon blockade with a large blockade radius, we use an electrically controlled photonic device that hosts electrically tunable dipolar exciton polaritons, as depicted in Fig. \ref{fig:Fig2}A. It consists of a strip waveguide (WG) with a 510~nm-thick core of Al$_{0.4}$Ga$_{0.6}$As with twelve 20~nm-thick GaAs quantum wells (QWs), on top of an Al$_{0.8}$Ga$_{0.2}$As clad and a conductive doped substrate. The lateral optical confinement is defined by a 5~$\mu$m-wide, 200~$\mu$m-long ITO strip. The calculated optical modal shape is plotted to the right of the WG drawing, while Fig. \ref{fig:Fig2}C shows a schematic cross-section of the WG \cite{supp_methods}. The WG supports both TE and TM  modes. Here we focus only on the TE-mode as it has the strongest coupling strength with the excitons \cite{walker2013exciton}. The TE-optical mode has a large overlap with 8 QWs \cite{rosenberg2016electrically}, leading to the formation of three exciton-polariton branches ($i=LP,MP,UP$) from the interaction of the confined TE-photon with the two excitonic modes X$_{LH}$, X$_{HH}$ \cite{rosenberg2016electrically}: $\ket{P_i}=\chi_{TE}^i(\beta)\ket{TE}+\chi_{X_{HH}}^i(\beta)\ket{X_{HH}}+\chi_{X_{LH}}^i(\beta)\ket{X_{LH}}$, where $\beta$ is the propagation wave-vector in the WG, and $E_i(\beta)$ is the energy of the $i$-th polariton mode. The excitonic fraction is therefore defined as $\left| 
\chi_{X}(\beta)\right|^{2} = |\chi_{X_{LH}}^i |^{2} + | \chi_{X_{HH}}^i |^{2}$. Fig. \ref{fig:Fig2}D shows the calculated $E_{LP,MP}(\beta)$, the two lowest polariton modes. In order to excite the WG-polaritons and measure them after they exit, a 1st-order Au grating couplers are fabricated at each end of the WG. Importantly, the WG is electrically gated, resulting in an applied electric field perpendicular to the QWs-plane. The electric field induces a quadratic Stark shift to the excitons, leading to a voltage-dependent red-shift of the resulting polariton modes (Fig. \ref{fig:Fig2}E). The field also polarizes the polaritons, inducing a voltage-dependent electric dipole moment \cite{rosenberg2016electrically,rosenberg2018strongly,liran2024electrically}, calculated in Fig. \ref{fig:Fig2}E. This induced dipole moment leads to enhanced dipolar interactions between the polaritons, with a magnitude dependent on the applied field, which was shown to be tunable over more than two orders of magnitude \cite{rosenberg2018strongly,liran2024electrically}.

The experimental setup is shown in Fig. \ref{fig:Fig2}A. We excite the WG-polaritons of the LP branch resonantly by a pulsed laser ($2.2$~ps) impinging on the input coupler with a wavelength and incident angle matching the desired energy and momenta on the LP dispersion. The temporal duration of the coupled pulse is $\tau_p=3.1-5.7$~ps limited by the LP linewidth  at the particular energy. After propagation in the WG, the output grating extracts the photons into the spatial, temporal, and spectral imaging setup or a Hanbury Brown-Twiss interferometer for photon correlation measurements. Cross-polarization and with spatial filtering are used to reject the resonant laser reflections.
 We also measure the group velocities of the LP, $v_{g}$, by measuring the input-output delays. $v_g$ are found to be smaller than the bare TE-mode velocity, $v_p$, and agree well with the calculated values extracted from the dispersion\cite{supp_methods}. As expected, $v_g\sim15-50$~$\mu$m/ps in the polariton energy range relevant for these experiments depend on the exciton fraction. From these measurements, the effective 2-polariton density $n=2/A$ can be extracted for each experiment, where the mode area $A$ is defined as $A=w\tau_p v_{g}$ and $w\simeq5$~$\mu$m is the lateral width of the TE WG-mode.

\begin{figure}[!thb] 
	\centering
	\includegraphics[scale=0.8]{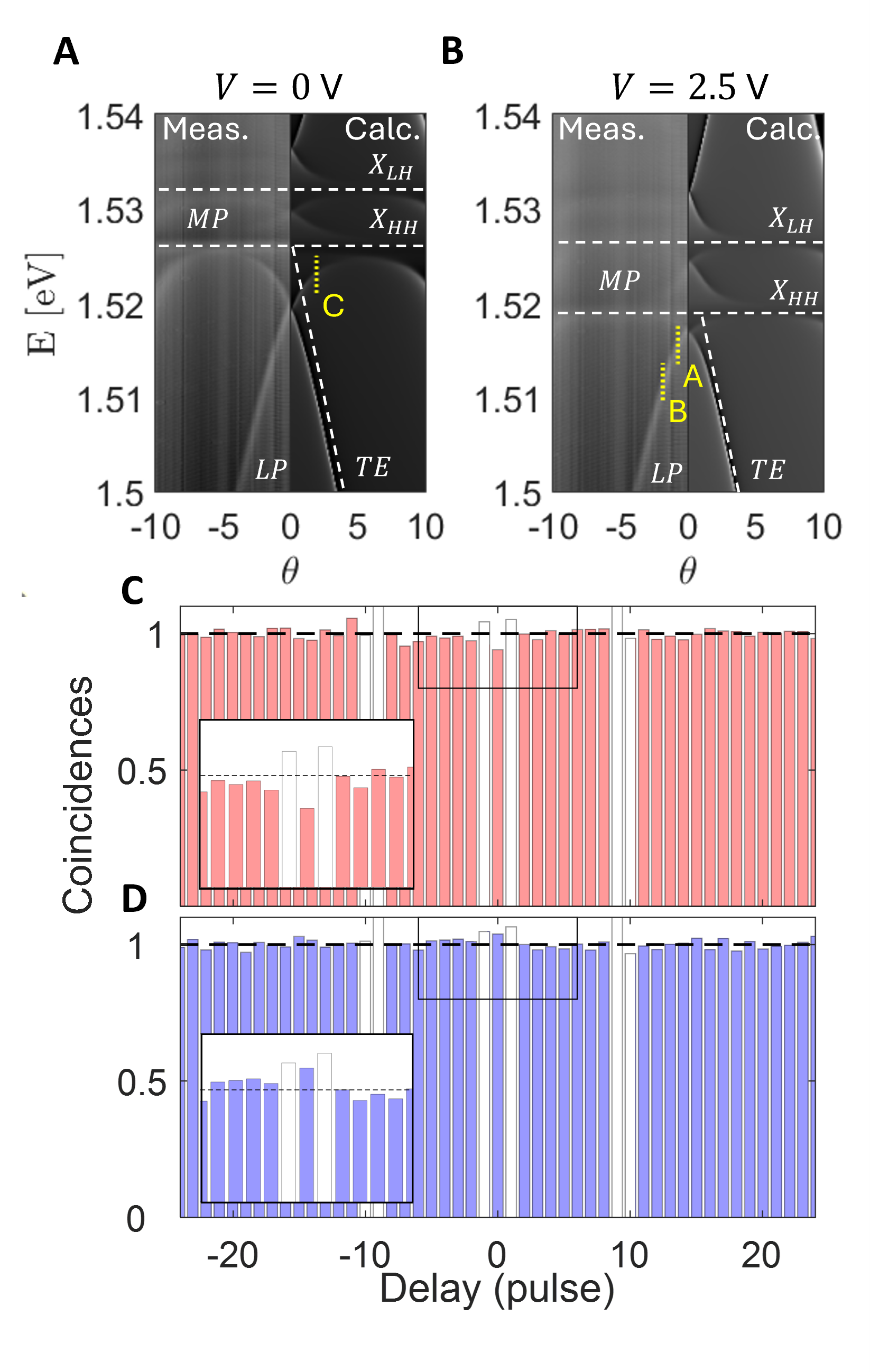} %

	\caption{\textbf{Photo-photon correlation measurements.}
		(\textbf{A-B})  Angular resolved differential reflection spectra (\(1-R/R_0\)) of the polaritons, at \(V = 0,2.5 V\). Left side is the measured spectrum, right side is calculated spectrum. The yellow lines on top of each spectrum, mark the laser injection range corresponding to the results of Fig. \ref{fig:correlation_detuning}. In white lines, we show the TE optical-mode, the exciton lines ($X_{hh},X_{lh}$), and the resulting lower and middle polariton branches (MP,LP).
        (\textbf{C-D}) Pulse-integrated coincidence counts for two laser detunings marked by the corresponding red and blue dots in Fig. \ref{fig:correlation_detuning}. The white pulses are discarded due to detectors crosstalk. The measurements yield an antibunching with $g^{(2)}(0)=0.94\pm 0.02$ for  $\Delta=-0.36$~meV (\textbf{C}), and bunching with $g^{(2)}(0)=1.030\pm 0.015$ for  $\Delta=+0.23$~meV (\textbf{D}). In the insets we show close-ups of the marked areas which show a clear deviation from the side peaks. }

	\label{fig:raw_correlations} %
\end{figure}

In the left panels of Fig. \ref{fig:raw_correlations}A,B  we show white-light differential reflection from the input grating for two voltages, $V=0,2.5$~V, as a function of the energy and incident angle. The LP modes of both right propagating and left propagating polaritons are clearly observed, with a significant Stark redshift of the whole $V=2.5$V dispersion. The right panels show the calculated differential reflection spectra, with a very good agreement to the measurements, allowing to extract the relevant parameters of the WG-polariton system \cite{supp_methods}.

To measure $g^{(2)}_0(\Delta)$, we sweep the laser energy in and out of the 1P transition at given incoming angles, corresponding to different propagation wave vectors $\beta$ and different excitonic fractions $|\chi_X(\beta)|^{2}$. The yellow dotted arrows depict the sweeps done in these measurements. In Fig. \ref{fig:raw_correlations}C,D we show normalized, pulse integrated coincidence counts which correspond to $g^{(2)}(\tau_m=mT)$, where $T=12.5$~ns is the time between pulses, and $m=0,\pm1,\pm2...$, for two distinct laser detunings and excitonic fractions on the biased WG. The black dashed lines correspond to a value of $g^{(2)}(\tau)=1$ extracted by averaging the values of the side peaks. We mark in white correlation peaks that are affected from an optical cross-talk inside the measurement system \cite{reinhard2012strongly,wood2019non,supp_methods}, which are discarded from our analysis. For a negative $\Delta=-0.36$~meV, a clear anti-bunching of $g^{(2)}_0= 0.94\pm 0.02$ is observed, which is a signature of a partial blockade, while for a positive $\Delta=+0.23$~meV, a bunching effect is measured with $g^{(2)}_0=1.030\pm 0.015$, pointing at an anti-blockade mechanism.

\begin{figure}[!thb] %
	\centering
	\includegraphics[scale=0.8]{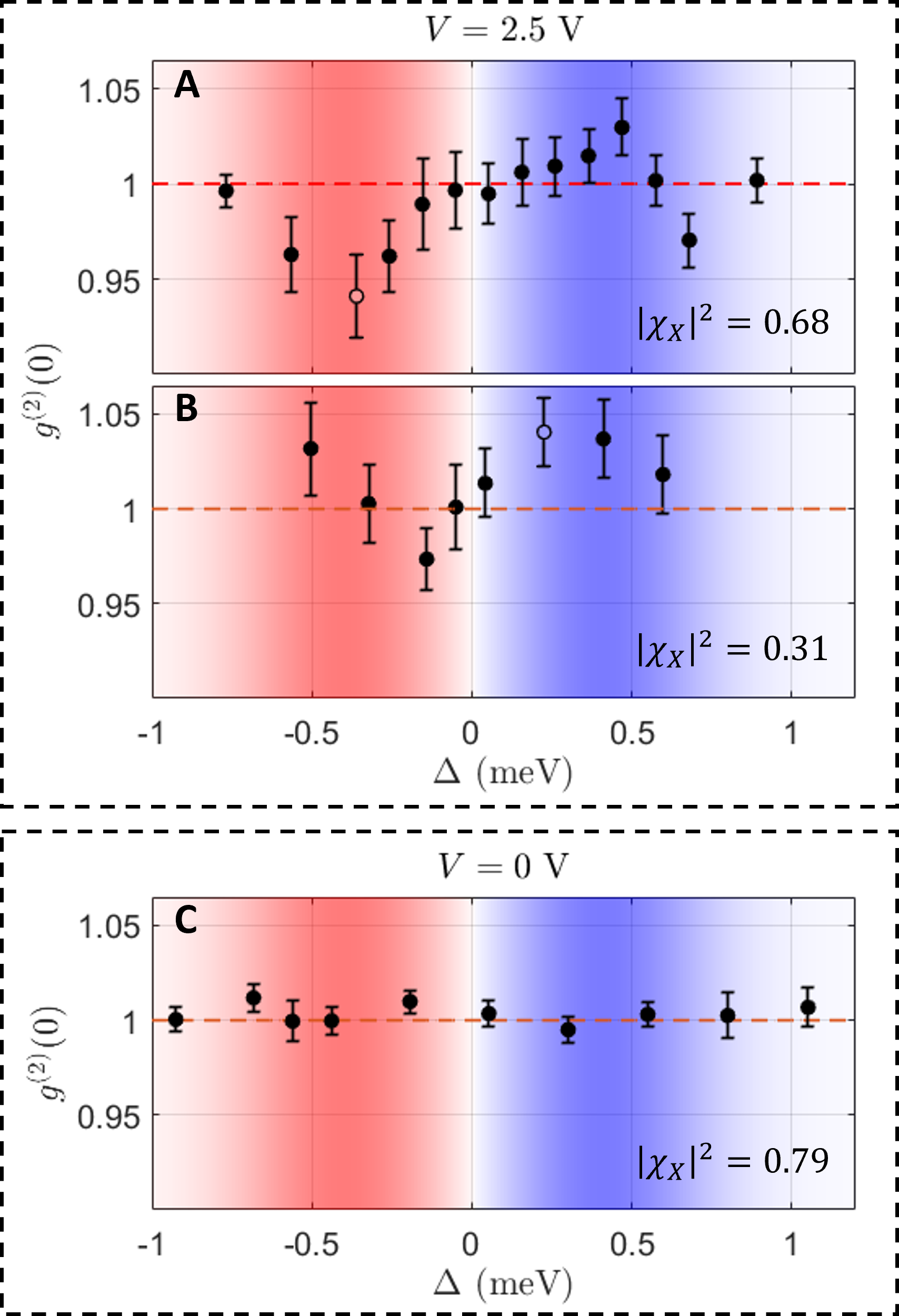} %

	\caption{\textbf{Voltage tunable dipolar Blockade and anti-blockade.}
		(\textbf{A}) $g^{(2)}(0)$ as a function of the laser detuning $\Delta$ and under an applied voltage $V=2.5$~V, displaying a clear anti-bunching and bunching, suggesting partial blockade and anti-blockade of dipolaritons.  The excitation marked  in Fig. \ref{fig:Fig2}(d) correspond to dipolaritons with a $68\%$ exciton fraction.
        (\textbf{B}) $g^{(2)}(0)$ under the same conditions as in (a) but for polaritons with a $31\%$ exciton fraction. The effect is reduced due to the larger velocity and mode area (corresponding to lower densities) of more photon-like dipolaritons. The colored dots in \textbf{A,B} correspond to the two points in Fig. \ref{fig:raw_correlations}
        (\textbf{C}) $g^{(2)}(0)$ for non-dipolar polaritons ($V=0$~V) showing a flat dependence with $g^{(2)}(0)\simeq1$. }
	\label{fig:correlation_detuning} 
\end{figure}

The expected blockade to anti-blockade transition of Fig. \ref{fig:conceptual}A with varying $\Delta$ is clearly observed in Fig. \ref{fig:correlation_detuning}A, showing the dependence of $g^{(2)}_0(\Delta)$ for dipolar polaritons under a voltage of $V=2.5$~V with an excitonic fraction of $68\%$. To verify that this effect is indeed due to the interaction between the excitonic part of the polaritons, Fig. \ref{fig:correlation_detuning}B display $g^{(2)}_0(\Delta)$ for dipolar-polaritons under the same voltage but with a lower excitonic fraction, indeed showing a decreased effect by roughly a factor of 2.
Finally, we show that the significant photon correlations arise from the strong interaction between electrically polarized polaritons: Fig. \ref{fig:correlation_detuning}C presents the correlations for non-dipolar polaritons ($V=0$~V). No correlations are observed, even for polaritons with a high excitonic fraction ($\sim 79\%$) that is expected to increase interactions both due to an increased matter fraction, and to a reduction of the the group velocity ($v_g\sim 17$~$\mu$m/ps) that in turn results in an increased 2-polariton density.

The measured two-photon correlation not only gives a qualitative signature of a polariton blockade. It was shown \cite{verger2006polariton,delteil2019towards} that the minimal value of $g^{(2)}_0(\Delta)$ is directly related to the interaction strength $U_{PP}$, via the relation:
\begin{equation}
    \kappa \frac{U_{PP}}{\gamma}\simeq1- g^{(2)}_{0,min},
\label{eq:g2interaction}
\end{equation}
where $\kappa$ is a numerical factor depending on the laser pulse length $\tau_p$, and on the linewidth $\gamma$. For a dipole-dipole interaction of polaritons having a dipole length $d$ we can write \cite{liran2024electrically}:
\begin{equation}
    U_{PP}\rightarrow U_{dd}=\frac{2g_{dd}(|\chi_{X}(\beta)|^2,V)}{A}.
\label{eq:interaction_const}
\end{equation}
For the data in Fig. \ref{fig:correlation_detuning}B (C), we set into Eqs. \ref{eq:g2interaction},\ref{eq:interaction_const} the following values: $\gamma\simeq215\mu$eV ($115\mu$eV), extracted from the linewidth \cite{liran2018fully,liran2024electrically}, $v_{g}=25.6$~$\mu$m/ps ($52.1$~$\mu$m/ps) which yield $A=385$~$\mu$m$^{2}$ ($1465$~$\mu$m$^{2}$) and $\kappa\simeq 0.61\pm0.01$ from numerical simulations \cite{supp_methods} respectively. We thus obtain an experimental value for $g_{dd} = 4.0 \pm 1.4$~meV$\mu$m$^{2}$  ($3.6\pm 2.2$~meV$\mu$m$^{2}$) for B (C). Remarkably, this result represents a two orders of magnitude increase over $g_{PP}$ of non-dipolar polaritons \cite{munoz2019emergence,delteil2019towards}, confirming our previous estimates from measurements of both density dependent energy shifts \cite{rosenberg2018strongly} and transmission\cite{liran2024electrically} of WG-dipolaritons in independent experiments on the same sample. This large magnitude of $U_{dd}$ can be even better appreciated by noting that here, values of $g^{(2)}_{0,min}$ similar to those in Refs. \cite{munoz2019emergence,delteil2019towards} are found, but for average 2-polariton densities two orders of magnitude smaller. This is largly a consequence of the extended range of \textit{dipolariton induced field screening effect}, which is the dominant term the interaction between two \textit{field-induced} dipoles \cite{liran2024electrically}.   

As mentioned above, an important parameter for evaluating the applicability of a blockade effect as a building block for a universal 2-photon gate is the blockade radius. From Eq. \ref{eq:blockade} we extract $R_{b}=3.4\pm0.6$~$\mu$m ($R_{b}=4.4\pm1.4$~$\mu$m) for our system (Supplementary Text). This yields $R_b\gg\lambda_{WG}$, where $\lambda_{WG}\simeq0.8\mu m/3.6=0.22$~$\mu$m is the optical wavelength inside the WG. It means that a full blockade is possible for realistic WG mode confinement areas which are larger than $\lambda_{WG}^2$, easily achieved with conventional WG structures. Remarkably, this value of $R_b$ is close to the values found in Rydberg atoms mentioned above, but in a solid-state based photonic chip. 

In our experiment, only a partial blockade is found, since $n=2/A \simeq 5\cdot {10}^{-3}$~$\mu$m$^{-2}$ which is much smaller than the blockade density $n_b\equiv \gamma/g_{dd}=0.05$~$\mu$m$^{-2}$, thus, $n/n_b\simeq0.1$. However, the full blockade condition $n>n_b$, can be readily reached by simple modifications of the current system. The blockade condition can be expressed as: 
\begin{equation} \label{eq:blockade}
    \frac{2}{A}>\frac{\gamma}{g_{dd}}.
\end{equation}
We set $A=v_{g}\tau_p w=v_{g}(\hbar/\gamma) w$, where $v_g\simeq v_p(1-|\chi_{X}|^{2})$ \cite{supp_methods}, and $g_{dd}\propto d |\chi_{X}|^{4}$ \cite{liran2024electrically}. Using these expressions and the experimentally extracted values, $g_{dd}^{ex}$, $d_{ex}$, and $|\chi_{X}^{ex}|^2$, Eq. \ref{eq:blockade} yields the following condition for a full photon-blockade (Supplementary Text): 
\begin{equation} \label{eq:constraint}
    \frac{w (1-|\chi_{X}|^{2})}{d |\chi_{X}|^{4}}<C_{\text{ex}}=\frac{2g_{dd}^{ex}(1-|\chi_{X}^{ex}|^{2})}{\hbar v_{g}^{ex} d_{ex}|\chi_{X}^{ex}|^{4}}\simeq 50.
\end{equation}
Our optical simulations \cite{supp_methods}  predict that a guided WG-polariton mode is supported by WGs as narrow as half a micron (having a mode of $w=0.28$~$\mu$m). With a dipole length of $d=8$nm (Fig. \ref{fig:Fig2}E), the blockade condition is therefore fulfilled for an exciton fraction $|\chi_{X}|^{2}>0.56$, which is easily realizable.

In summary, we have demonstrated an on-chip semiconductor-based device, having WG-dipolariton interactions with a continuously tunable strength up to a photon-blockade radius comparable with that of Rydberg atoms, and much larger than the optical wavelength. 
This, and our recently demonstrated WG-dipolariton transistor device~\cite{liran2024electrically}, both 
allow precise electrical control, have tiny active footprints of a few micrometers in length,
and are readily integrated with other WG-optical elements such as couplers and channels. 
These devices can be switched electrically with a GHz bandwidth and a record low power consumption of less than 3fJ per gate operation \cite{Dror_unpublished}, which are all crucial for scaling up to large circuits. 
These advances show that WG dipolar polaritons provide a new platform for scalable, ultrafast-reconfigurable quantum photonic circuits that uniquely combines strong two-photon nonlinearities, scalability and ready integration with both optics and electronics.

\clearpage %
\bibliography{references} 
\bibliographystyle{style}

%


\section*{Acknowledgments}

The authors wish to thank Yaron Bromberg for a critical reading of the manuscript.

\paragraph*{Funding:}
R.R., Y.O. and D.L. acknowledge the support from the Israeli Science Foundation Grant 1087/22, and from the NSF-BSF Grant 2019737. H.D. acknowledges the support of the National Science Foundation under grant DMR 2004287, the
Army Research Office under grant W911NF-17-1-0312, and the Air Force Office of Scientific Research under grant FA2386-21-1-4066, and the Gordon and Betty Moore Foundation under
grant N031710. This research is funded in part by the Gordon and Betty Moore Foundation’s
EPiQS Initiative, Grant GBMF9615 to L. N. Pfeiffer, and by the National Science Foundation
MRSEC grant DMR 2011750 to Princeton University
\paragraph*{Author contributions:}
Y.O., D.L., and R.R. conceptualized the project. K.B. and L.P. grew the sample. D.L. and R.R. designed the device. D.L. performed the device fabrication. D.L. Y.O. and R.R. designed the experimental setup. Y.O. and D.L. performed the measurements and the data analysis under the supervision of R.R., with consultation with H.D. Y.O. performed numerical simulations with input from R.R. The manuscript was written by Y.O., D.L., H.D., and R.R.

\paragraph*{Competing interests:}
There are no competing interests to declare.


\subsection*{Supplementary materials}
Materials and Methods\\
Supplementary Text\\
Figs. S1 to S7\\
References \textit{(7-\arabic{enumiv})}\\ %


\newpage


\renewcommand{\thefigure}{S\arabic{figure}}
\renewcommand{\thetable}{S\arabic{table}}
\renewcommand{\theequation}{S\arabic{equation}}
\renewcommand{\thepage}{S\arabic{page}}
\setcounter{figure}{0}
\setcounter{table}{0}
\setcounter{equation}{0}
\setcounter{page}{1} %


\begin{center}
\section*{Supplementary Materials for\\ \scititle}

	Yoad~Ordan$^{1\dagger}$,
	Dror~Liran$^{1\dagger}$,
        Kirk~W.~Baldwin$^{2}$,
        Loren~Pfeiffer$^{2}$,
        Hui~Deng$^{3}$,\and
	Ronen~Rapaport$^{1\ast}$\\
	\small$^{1}$Racah Institute of Physics, The Hebrew University of Jerusalem, Jerusalem 9190401, Israel.\\
        \small$^{2}$Department of Electrical Engineering, Princeton University, Princeton, NJ 08544, USA.\\
	\small$^{3}$University of Michigan, Ann Arbor, MI 48109, USA.\\
	\small$^\ast$Corresponding author. Email: ronenr@phys.huji.ac.il\\
	\small$^\dagger$These authors contributed equally to this work.
\end{center}

\subsubsection*{This PDF file includes:}
Materials and Methods\\
Supplementary Text\\
Figures S1 to S7\\

\newpage


\subsection*{Materials and Methods}

\subsubsection*{Sample design and structure}
\begin{figure}[!thb]
    \centering
    \includegraphics[width=0.6\linewidth]{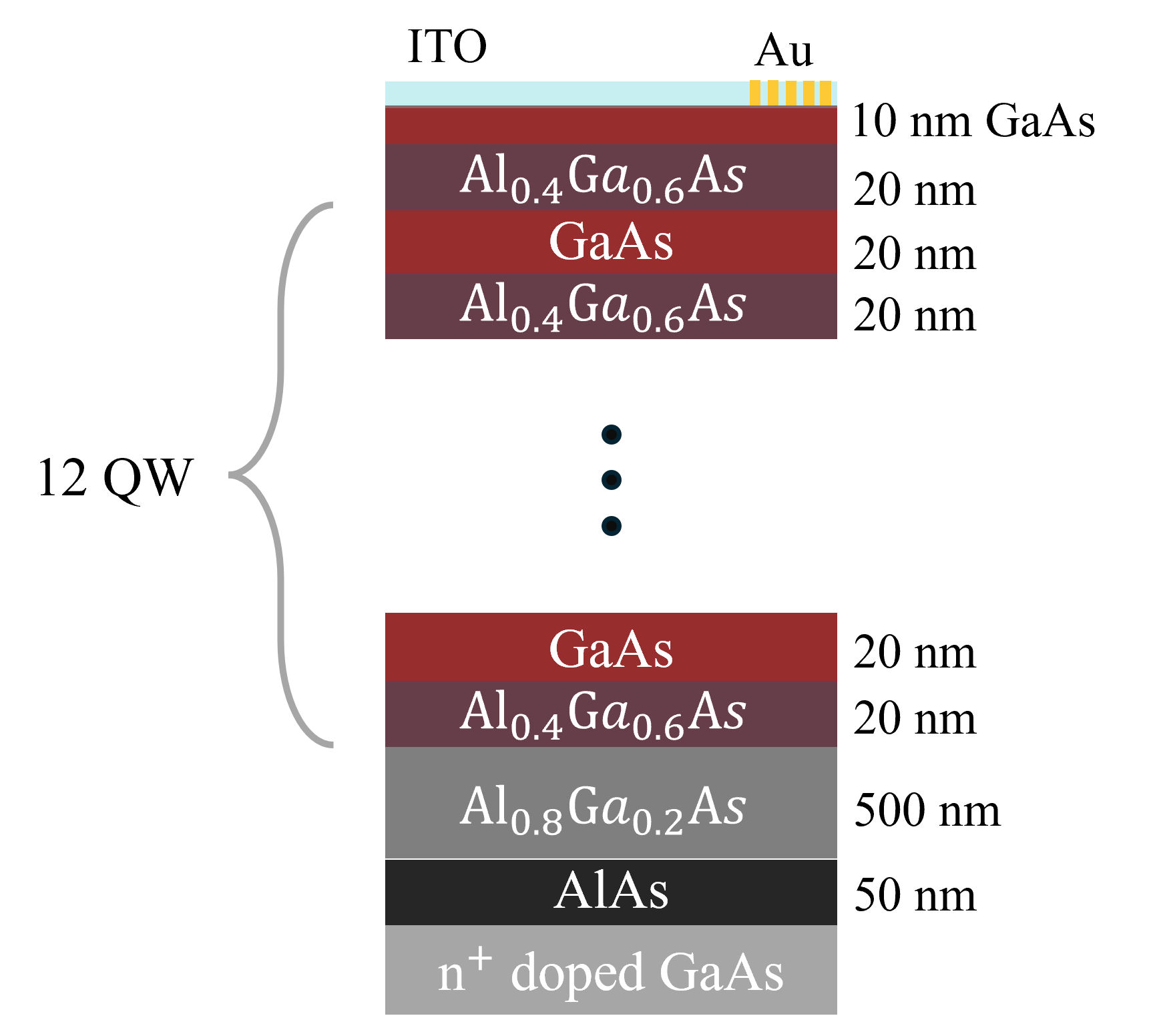}
    \caption{\textbf{The sample structure.} A sketch of the different layers including the clad, the core which consists of twelve quantum wells and the ITO strip along with the gold gratings.}
    \label{fig:supp_sample_diagram}
\end{figure}

The sample is based on a multi-QW structure that also serves as a slab WG for photons, and is identical to the one in Ref. \cite{liran2024electrically}. The sample is grown on n+ doped GaAs substrate that also serves as the bottom electrode for electrical gating. The bottom clad part is composed of 50~nm layer of AlAs and 500~nm layer of Al$_{0.8}$Ga$_{0.2}$As which both have lower refractive index compared to the upper core and therefore allow for confined optical modes in the vertical direction \cite{rosenberg2016electrically,rosenberg2018strongly}. The core is made of twelve, 20~nm thick GaAs QWs embedded with equal separation within a 500~nm layer of Al$_{0.6}$Ga$_{0.4}$As. A thin (10~nm) top layer of GaAs is placed to prevent oxidation. The sample is grown using molecular-beam epitaxy (MBE) and the structure is shown in Fig. \ref{fig:supp_sample_diagram}. The photon coupling gratings on the sample are made by evaporating gold with a period of 240~nm which is chosen to diffract the optical mode of the WG perpendicular to the sample for it to be collected by the optical setup. 
The gratings are achieved by coating the sample with a 120~nm layer of PMMA (950A2) which is patterned by using the ELS-G100 by ELIONIX (500~pA/ 100~kV). The PMMA is then exposed to a dose of 1500~$\mu$C/cm$^{2}$ to write the gratings. Next, the PMMA is developed at $-$5$^{\circ}$C for one minute. Then 10~nm Ti and 50~nm Au layers are deposited and the unwanted regions are discarded using lift-off process. Then we deposit the 50~nm thick, 5~$\mu$m wide layer of Indium-Tin-Oxide (ITO) and pattern it using laser-lithography using a AZ1505 photo-resist. Finally the now patterned layer of ITO is wet-etched using 1:1 ratio, H$_{2}$O:HCL. The ITO strip completes the WG geometry and allows for confined mode also in the lateral direction. The ITO has both electrical conductivity to allow for gating and a higher than air refractive index to confine the optical modes (along with a relatively low absorption at the relevant wavelengths).For the wire bonding to the sample we also evaporate 100~nm thick Au contacts. 

\subsubsection*{Experimental setup and measurements}
\begin{figure}[!thb]
    \centering
    \includegraphics[width=0.9\linewidth]{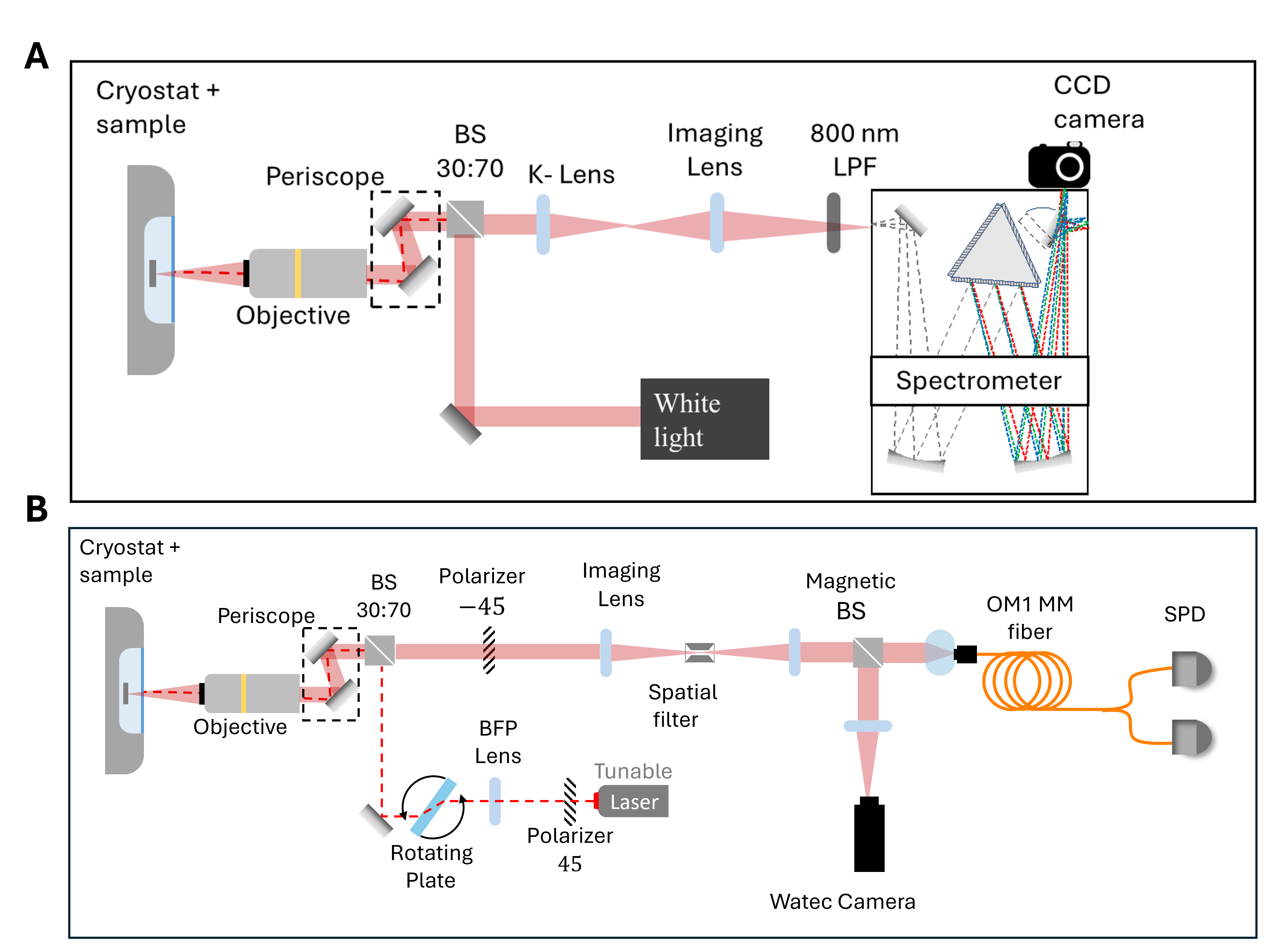}
    \caption{\textbf{The experimental configurations.} 
    (\textbf{A}) A white light source is focused on the gratings with wide angle range. The resulting reflection spectrum of the polaritons is coupled to the optical system and its Fourier plane is imaged onto a spectrometer. (\textbf{B}) The resonant excitation scheme showing the BPF which is used to tune the excitation wavelength and the rotating plate which is used to vary the angle of incidence. The polaritonic emission is imaged, filtered with cross-polarization and spatial filtering and is then coupled into a multimode fiber which is connected to a fiber-based HBT setup.}
    \label{fig:supp_system}
\end{figure}

To measure the polaritonic dispersion and find the relation between the angle incident angle of the absorbed photons and the wavelength (WL) (Fig. \ref{fig:Fig2}), we perform a white light absorption measurement: we focus a white light source with wide range of angles on the input grating. The light reflection is coupled into the optical Fourier spectral imaging system that yields a WL Vs. angle image as is shown in Fig. \ref{fig:supp_system}A. The result is normalized to a reflection measurement  on  a flat gold pad that has no polaritonic absorption. \\
Once we have extracted a dispersion, we can excite any point on the dispersion by tuning the incident laser energy and angle to match that of the required point on the white light image. 

For the resonant polariton excitation we use a 80~Mhz, 100~fs Ti:Sapphire laser (Spectra-Physics Mai-Tai). To fit the laser spectrum to be similar to the spectral width of the polaritonic mode, measured from the absorption spectrum, we pass the laser through an ultra narrow bandpass filter with a central WL of 828~nm and FWHM of 0.5~nm (Alluxa 828-0.5). In this way we can tune the central wavelength of the laser by simply by rotating the filter. Next, to have a fine tuning of the angle of incidence, we focus the laser on the back focal plane of the imaging objective such that it will arrive at the sample with a well defined angle (while increasing the laser spot size to roughly $20$~$\mu$m). To control this angle a rotating glass plate is placed on the laser path as shown in Fig. \ref{fig:supp_system}B. Rotating the glass causes a displacement of the laser path (Snell's law) which will transform to a shift in the angle of incidence by the objective. 

\begin{figure}[!thb]
    \centering
    \includegraphics[width=0.85\linewidth]{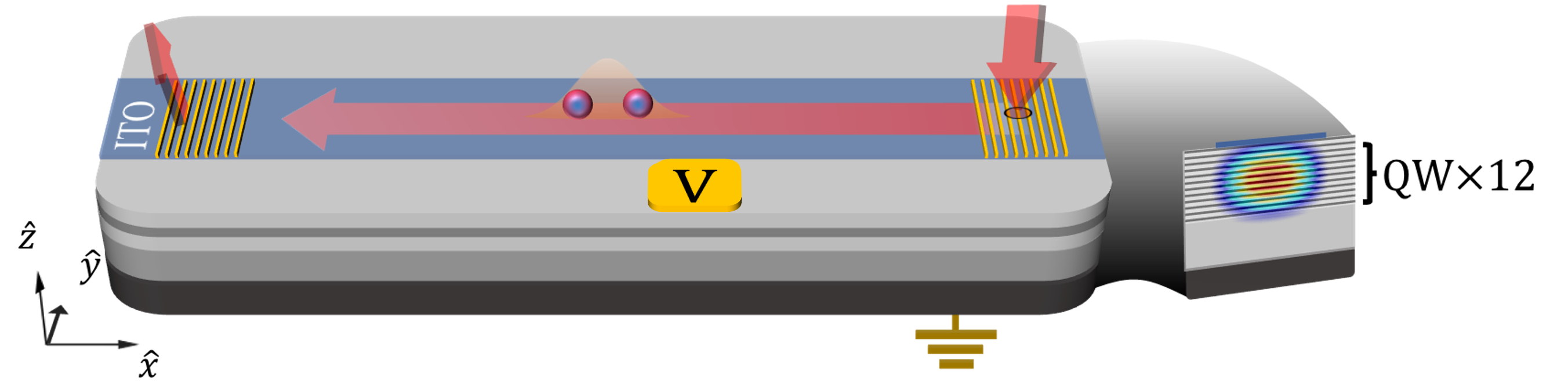}
    \caption{\textbf{Polariton propagation in the WG channel.} When the input grating is excited with a laser with energy and angle that matches that of a polariton mode, the polaritons propagate along the channel. Once the packet reaches the exit grating, the mode is coupled out resulting in a photon that matches the polariton's energy and angle of emission that is uniquely defined by the momenta of the polaritons in the channel.}
    \label{fig:supp_channel}
\end{figure}

Once a polaritonic pulse is excited in the grating, it propagates in the 200~$\mu$m long channel to the second grating where the polaritonic mode  couples out as shown in Fig. \ref{fig:supp_channel}. To filter the laser and obtain a high SNR between the polaritonic signal and the background laser, we use a cross-polarization scheme in which we excite the TE-mode (90$^{\circ}$) with a 45$^{\circ}$ laser (that reduces the coupling efficiency by half) and then place an orthogonal polarizer in the emission path such that the signal is reduced by half while the laser is suppressed by the extinction ratio ($\sim 10^{6}$) of the polarizers. To further filter the laser noise we use a movable spatial slit filter in an image plane of the device, allowing for a selective choice of the region of interest. Because the signal is emitted from the second grating (Fig. \ref{fig:supp_channel}) we place the slit to match the output grating such that only light that originated from that point on the sample will pass through. To verify the slit's position a beam splitter and a camera is placed. Once the slit has been placed the beam splitter is removed and the entirety of the signal is coupled in to a multimode fiber (OM2, Graded-index, $62.5$~$\mu$m core). The fiber is then connected to a fiber-based HBT setup which consists of a fiber-splitter (one input port and two output ports with 50:50 splitting in intensity) and two fiber-connected single photon detectors (Excelitas SPCM-AQRH-14-FC). The detection events are then registered and correlated on a time-correlator (Swabian TimeTagger 20).

\subsubsection*{Calculation of the WG modes and the polariton dispersion}
The optical modes in the strip waveguide can be estimated analytically or fully solved numerically. For an analytical estimation, we assume that the z and y directions of the optical modes are decoupled; thus, the full two-dimensional mode is composed of the one-dimensional solutions for the two directions, each is given by the solution of:
\begin{equation}
    \dv[2]{E_i(j)}{j}+\qty(k_0^2n(j)^2-\beta^2)E_i(j)=0 \label{eq:WGMode}
\end{equation}
here $i\perp j$ are the z,y axes.
The mode that satisfies Eq.\ref{eq:WGMode} results in a transcendental equation, which is then solved numerically.
To solve the mode in the $\hat{y}$ direction, we first calculate the modes in the $\hat{z}$ direction in the presence of an ITO strip and without it. From the modes, we can extract the effective index of refraction with and without ITO. Finally, we use the effective indices of refraction and solve for the structure with the index calculated for the ITO strip, sandwiched by the index calculated without the ITO strip, to find the modes in the $\hat{y}$ direction. In Fig. \ref{fig:Fig2}B, we plot the mode profile in the $\hat{y}$ direction calculated this way.

\begin{figure}[!thb]
    \centering
    \includegraphics[width=0.9\linewidth]{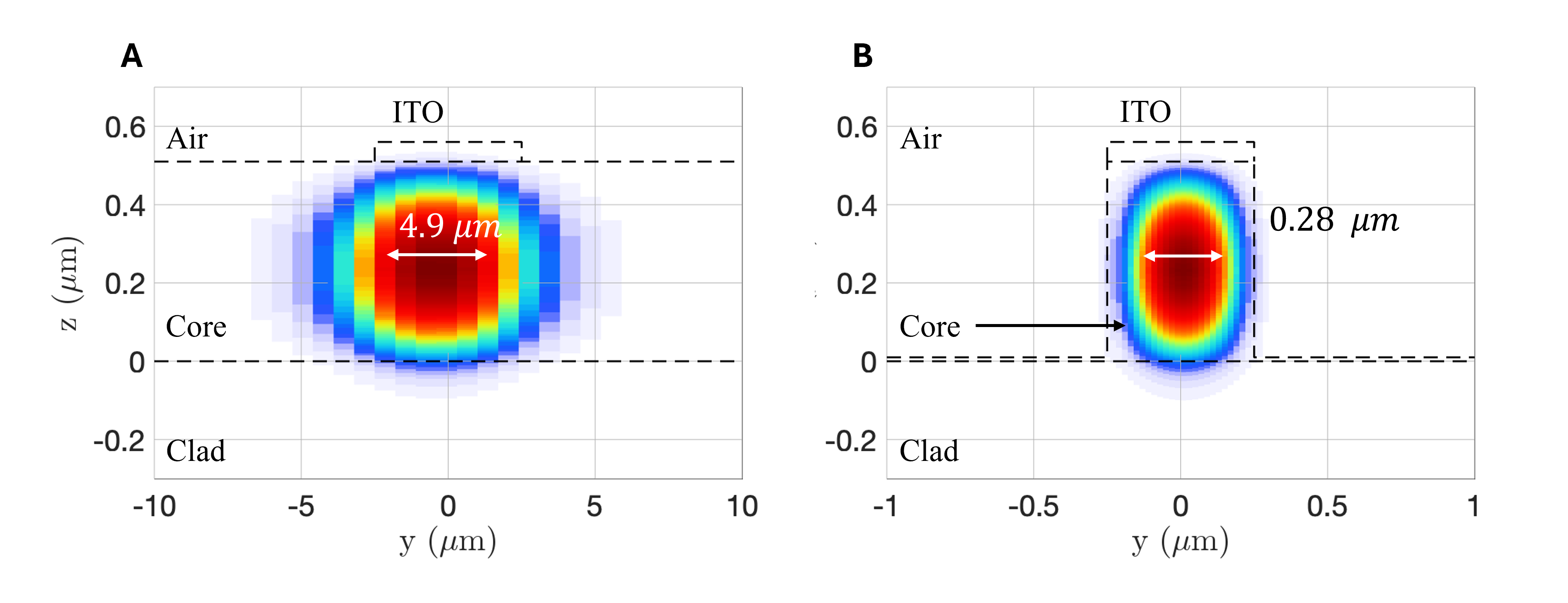}
    \caption{\textbf{Calculated WG modes.} The WG modes calculated in an FDE solver along with the physical dimensions of the device for (\textbf{A}) a 5~$\mu$m device showing a mode width a FWHM of $4.9$~$\mu$m and for (\textbf{B}) a $0.5$~$\mu$m wide device which contains a mode width a FWHM of $0.28$~$\mu$m.}
    \label{fig:supp_mode}
\end{figure}

To solve the mode accurately, a full numerical solution of the Maxwell equations in the plane without sources is required; we perform it with a "Lumerical" Finite-difference-element (FDE) solver. In Fig. \ref{fig:supp_mode}, we plot the calculated modes for a strip waveguide with 5 $\mu m$ wide ITO Layer and for a 0.5~$\mu$m wide waveguide with etched sidewalls.

The theoretical differential reflection part in Fig. \ref{fig:raw_correlations}A,B was calculated using a home-built RCWA (Rigorous coupled wave analysis) solver \cite{rosenberg2016electrically,Schwarz2012Gratings}.
The parameters for the Exciton and Exciton-photon coupling used are $E_{hh} = 812 nm,\, E_{lh} =809.3 nm,\,  \Omega_{hh} = 6.4 meV,\, \Omega_{lh} = 4.5 meV$ for V=0 and $E_{hh} = 817.8 nm,\, E_{lh} = 811.7 nm,\,  \Omega_{hh} = 5.4 meV,\, \Omega_{lh} = 3.7 meV$ for V = 2.5 V.

\subsubsection*{$g^{(2)}(0)$ and optical cross-talk}

\begin{figure}[!thb]
    \centering
    \includegraphics[width=0.8\linewidth]{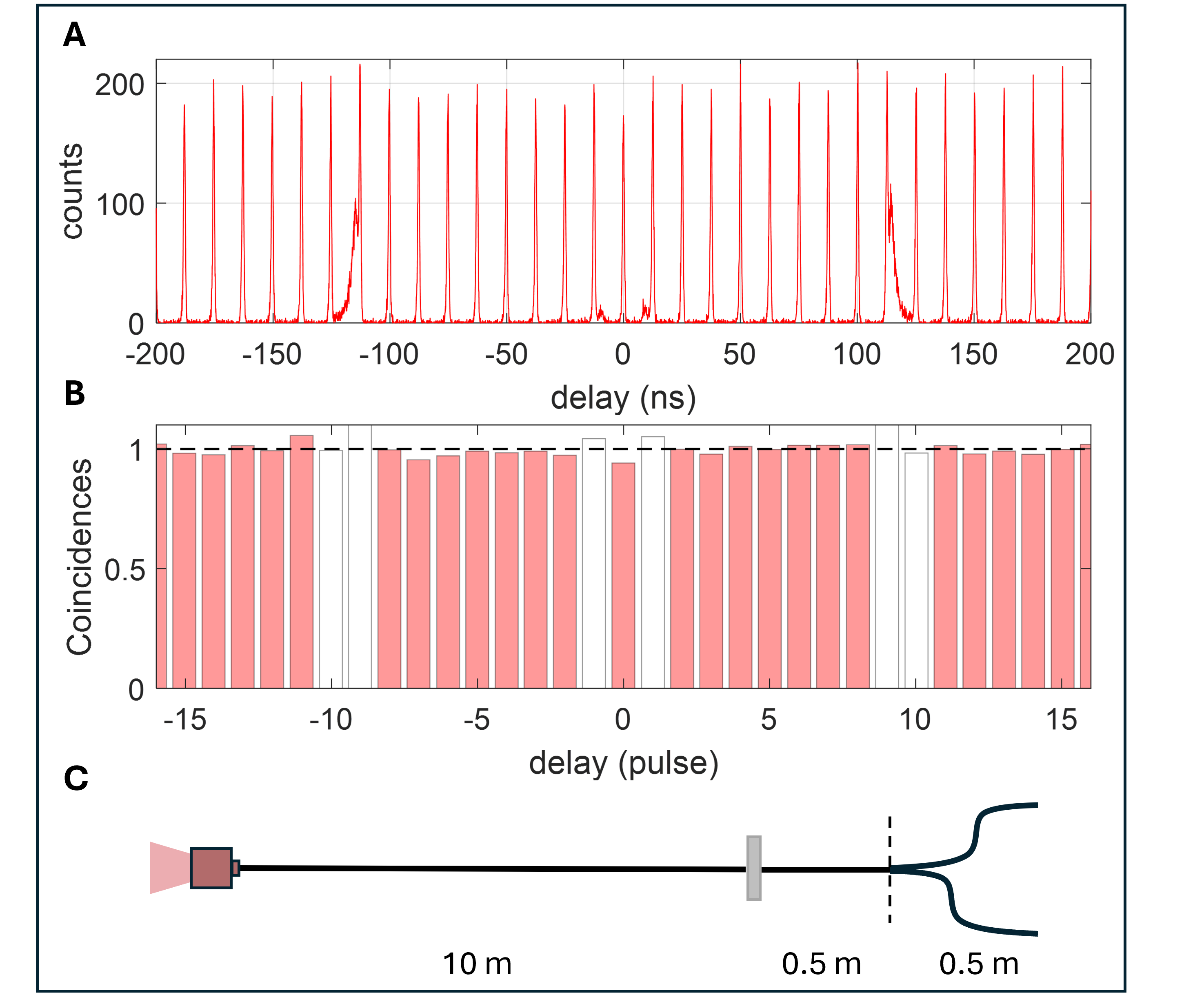}
    \caption{\textbf{Coincidence counts analysis.} 
    (\textbf{A}) raw coincidence counts for different delays, each point correspond to a 87~ps bin.
    (\textbf{B}) integrated counts for each pulse showing the fluctuations in the side peaks values and the average value in a black dashed line. The bars which are affected from the optical cross-talk are marked in white and are discarded from the calculation. (\textbf{C}) The fiber-based setup showing the objective that couples in the emission into a 10~m long fiber which is connected to a fiber splitter with 0.5~m length input and output arms. The interface between the two fibers and between the fiber and the air (in the coupling point) have residual reflectance. }
    \label{fig:supp_corr_cross}
\end{figure}

The HBT measurement results in coincidence counts of all the detection events in the two detectors of the HBT setup with the delay between them as is shown in Fig. \ref{fig:supp_corr_cross}A. The width of each peak is roughly $700$ ps and is limited by the jitter of the detection system and not by the  poalriton pulse length (which is measured to be $<6$~ps). Therefore correlations shorter than this time cannot be resolved but the whole inter-pulse correlation can be deduced through integration of each pulse and the comparison of the integrated values as shown in Fig. \ref{fig:supp_corr_cross}B. Since the polariton pulse is much shorter than the time between pulses (12.5~ns), any coincidence counts which originated from two different pulses are completely independent which correspond to $g^{(2)}(\tau\neq0)=1$. Therefore the non-zero-delay coincidence counts can be used to normalize the central peak ($\tau=0$) and to obtain $g^{2}_{0}$. 

In principal all peaks for $\tau\neq0$ should be identical in their value, but the finite measurement time leads to statistical fluctuations that allow us to obtain an uncertainty to our normalized value of $g^{2}_{0}$.
The uncertainty in each of the values can be estimated through the fluctuation (standard deviation) of the side peaks, $\sigma_{S}$. The uncertainty of the central peak correspond to the fluctuation directly, $\sigma_{S}$, where for the average value of the side peak, $S$, it has a factor of $\sigma_{S}/\sqrt{N}$.
where $N$ is the number of side peaks. Defining $A(\tau=mT)$ to be the bin value for each delay, the central and side peaks values ($C,S$) are therefore:
\begin{align}
    C&\pm \Delta C =A(\tau=0) \pm\sigma_{S} \\
    S&\pm \Delta S = \frac{1}{N} \left( \sum_{m\neq 0} A(mT) \right) \pm \frac{\sigma_{S}}{\sqrt{N}}
\end{align}
Finally the value of $g^{2}_{0}$ is given by:
\begin{equation}
    g^{2}_{0} = \frac{C}{S} \pm \sqrt{\left( \frac{\Delta C}{S}\right)^2 + \left( \frac{C}{S^{2}} \Delta S\right)^{2}}.
\end{equation}

In Fig. \ref{fig:supp_corr_cross}A one can also see peculiar counts at roughly $9.5,114$~ns. We associate these peaks to an optical cross-talk between the two avalanche photo-diodes (APD) used in our experiment, which was also observed  in previous experiments \cite{munoz2019emergence,wood2019non}. In these APD detectors a single photon creates an avalanche of exciton-hole pairs that allow for a relatively high voltage/current response to a single photon. However, some of these pairs can recombine radiatively and be emitted back into the optical system. An existence of a reflective element somewhere in the system (such as a filter or an interface between two mediums) can reflect these photons back into the second detector and trigger a second detection event which will be after a time defined by the optical length the photons did from one detector to the other. Considering a propagation rate of roughly $5$~ns/m in the fibers, maps the temporal delay into a physical length which correspond roughly to $\sim2,23$~m. An analysis of the fiber setup in our system, which is shown in Fig. \ref{fig:supp_corr_cross}C, shows that there are two strong candidates for a reflection in our system, the coupling point where the light is coupled into the fiber from the air, and the interface between the two connected fibers. Considering the length of all the components, a photon that is reflected from either one of them back into the second detector will pass through a distance of $\sim 2,22$~m which further confirms the optical cross-talk effect.

\subsubsection*{Polariton velocity}
\begin{figure}[!thb]
    \centering
    \includegraphics[width=0.7\linewidth]{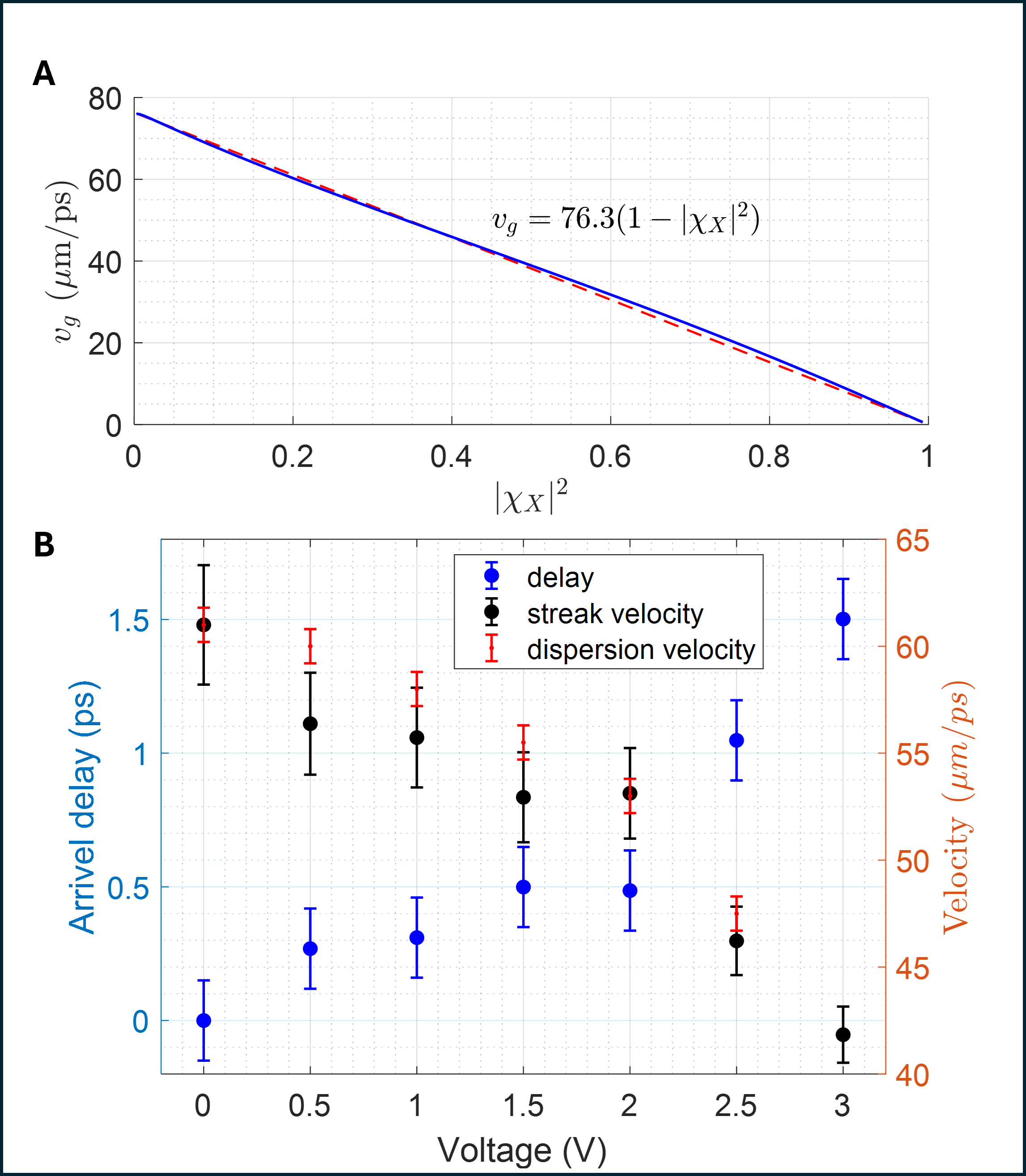}
    \caption{\textbf{ Polariton velocity analysis.}
    (\textbf{A}) The polariton group velocity extracted from the calculated dispersion as a fraction of the excitonic fraction showing a linear decrease of the velocity with the excitonic fraction.
    (\textbf{B}) Polaritons arrival times for different voltages (corresponding to different excitonic fractions) compared to the case of $V=0$~V showing an increased delay with increasing voltage (blue,left). The calculated group velocity is shown in red (right). The extracted group velocities from the streak measurements are shown in black, with a good agreement between the experiment and modeling.}
    \label{fig:supp_velocity}
\end{figure}

To extract the polariton pulse area $A=w\tau_p v_g$ (and thus the 2-polariton density), a knowledge of the polariton group velocity $v_g$ is required. The group velocity of a mode can be extracted from the dispersion (energy vs. momentum), $E(\beta)$, using the equation:
\begin{equation}
    v_{g}  = \frac{1}{\hbar}\frac{\partial E}{\partial \beta},
\end{equation}
which, when presented as a function of the excitonic fraction, shows linear dependence ($v_{g}\propto 1-|\chi_{X}(\beta)|^{2}$) as shown in Fig. \ref{fig:supp_velocity}A. To confirm this dependence we measure the arrival times of a polaritonic pulse after resonant excitation with a defined energy on a streak camera (Optronis OptosScope), and extract the group velocity from the delay between the input and output signal. This is done for a range of applied voltages. Due to the voltage induced redshift of the exciton and thus the LP mode, increasing the voltage results in an increased excitonic fraction of the propagating polariton. In parallel, the dispersion was calculated for each of these different voltages.  From each calculated dispersion, $v_g$ was extracted at the relevant excitation energy ($1.5122$~eV in our case). The comparison between the calculation and measurement is presented in red in Fig. \ref{fig:supp_velocity}B (right). We note that due to the arbitrary start of the streak camera compared to the polaritonic pulse, the absolute value of the time of arrival is not well defined. However, the difference in the arrival times for the different voltages can be used to extract information on the velocities and is shown in Fig. \ref{fig:supp_velocity}B in blue (left). Using the knowledge of the channel length, $x=200$~$\mu$m, and a known velocity, $v$, for one of the points, which we take from the dispersion calculation for $V=0$~V, one can calculate the velocity of the rest of the points: defining the arrival delay between the calculated point and the $V=0$~V point, $\Delta t$, the velocity of the calculated point, $v'$, is given by:
\begin{align}
    \Delta t &= x (\frac{1}{v'}- \frac{1}{v}) \\
    \frac{1}{v'} &= \frac{\Delta t}{x} + \frac{1}{v}.
\end{align}
The results of this method are presented in Fig. \ref{fig:supp_velocity}B in black (right) showing a very good  agreement with $v_g$ calculated from the  dispersion. 


\subsubsection*{Blockade model}

An alternative way we use to relate the correlation measurements results to the fundamental interaction strength is by a numerical solution of the polariton blockade model based on Ref. \cite{verger2006polariton}. Due to the large energy separation between the LP, MP and the UP branches we neglect the higher modes modes and keep only the the LP mode operators as presented in \cite{delteil2019towards} to obtain the anharmonic blockade model \cite{zubizarreta2020conventional} in the rotating-wave approximation which contains the LP ladder operators, $a,a^{\dagger}$, the polariton energy, $E_{p}$, the laser frequency, $\omega_{L}$, the polariton-polariton interaction, $U_{dd}$, and the laser strength and temporal shape, $\mathcal{F}(t)$:
\begin{equation}
    H = (E_{\text{p}}-\hbar \omega_{L}) a^{\dagger}a + \frac{U_{dd}}{2} a^{\dagger}a^{\dagger}aa+\hbar \mathcal{F}(t) (a+a^\dagger).
\end{equation}
To accurately describe the blockade meachnsim in our system, the model should also include the linewidth of the polaritonic mode which is done by introducing a polaritonic loss, $\gamma_{p}$, and evolving the state in time using the Lindblad master equation for the density matrix, $\rho$:
\begin{equation}
    \frac{\partial \rho}{\partial t} = \frac{i}{\hbar} \left[ \rho, H \right] +\gamma_{p} \left( a\rho a^{\dagger} -\frac{1}{2} \left( a^{\dagger}a\rho+\rho a^{\dagger}a \right) \right).
\end{equation}
After solving for $\rho(t)$ the (normalized) second-order correlation function, $G^{(2)}(t,t')$ ($g^{(2)}(t,t')$) can be written using the time evolution operator from $t'$ to $t$, $\mathcal{U}_{t,t'}$:
\begin{equation} \label{eq:g2_theory}
    g^{(2)}(t,t')= \frac{Tr \left( a\mathcal{U}_{t,t'} \left[ a\rho(t')a^{\dagger} \right] a^{\dagger} \right)}{Tr \left( a\rho(t)a^{\dagger} \right) Tr \left( a\rho(t')a^{\dagger} \right)} = \frac{G^{(2)}(t,t')}{Tr \left( a\rho(t)a^{\dagger} \right) Tr \left( a\rho(t')a^{\dagger} \right)}.
\end{equation}
To obtain the zero-delay value, $g^{(2)}_{0}$ in principle, one has to calculate Eq. \ref{eq:g2_theory} for $t=t'$. However our experiments does not have the required temporal resolution to observe such effects which therefore requires to use of pulsed excitation and to measure an averaged correlation over the whole pulse \cite{munoz2019emergence}. The right quantity to look at the simulation is therefore the integrated correlation over a period encapsulating one pulse, $\left[ -T,T\right]$, where $N(t) = Tr \left( a \rho a^{\dagger} \right)$:
\begin{equation}
    g^{(2)}_0 = \frac{2\int_{-T}^{T} dt_{1} \int_{t_{1}}^{T} dt_{2} G^{(2)}(t_{1},t_{2})}{\int_{-T}^{T}N(t_1) dt_{1} \int_{-T}^{T} N(t_2) dt_{2}}.
\end{equation}
Next, this quantity can be analyzed for different detunings $\Delta$, resulting in $g^{(2)}_0(\Delta)$ curves as shown in Fig. \ref{fig:conceptual}B. We note that the energy position of the maximum and minimum of the zero-delay correlation is mostly dependent on $\gamma_{p}$. The magnitudes of the dip and peak on the other hand depend on the ratio $U_{dd}/\gamma_{p}$. 

\begin{figure}
    \centering
    \includegraphics[width=0.8\linewidth]{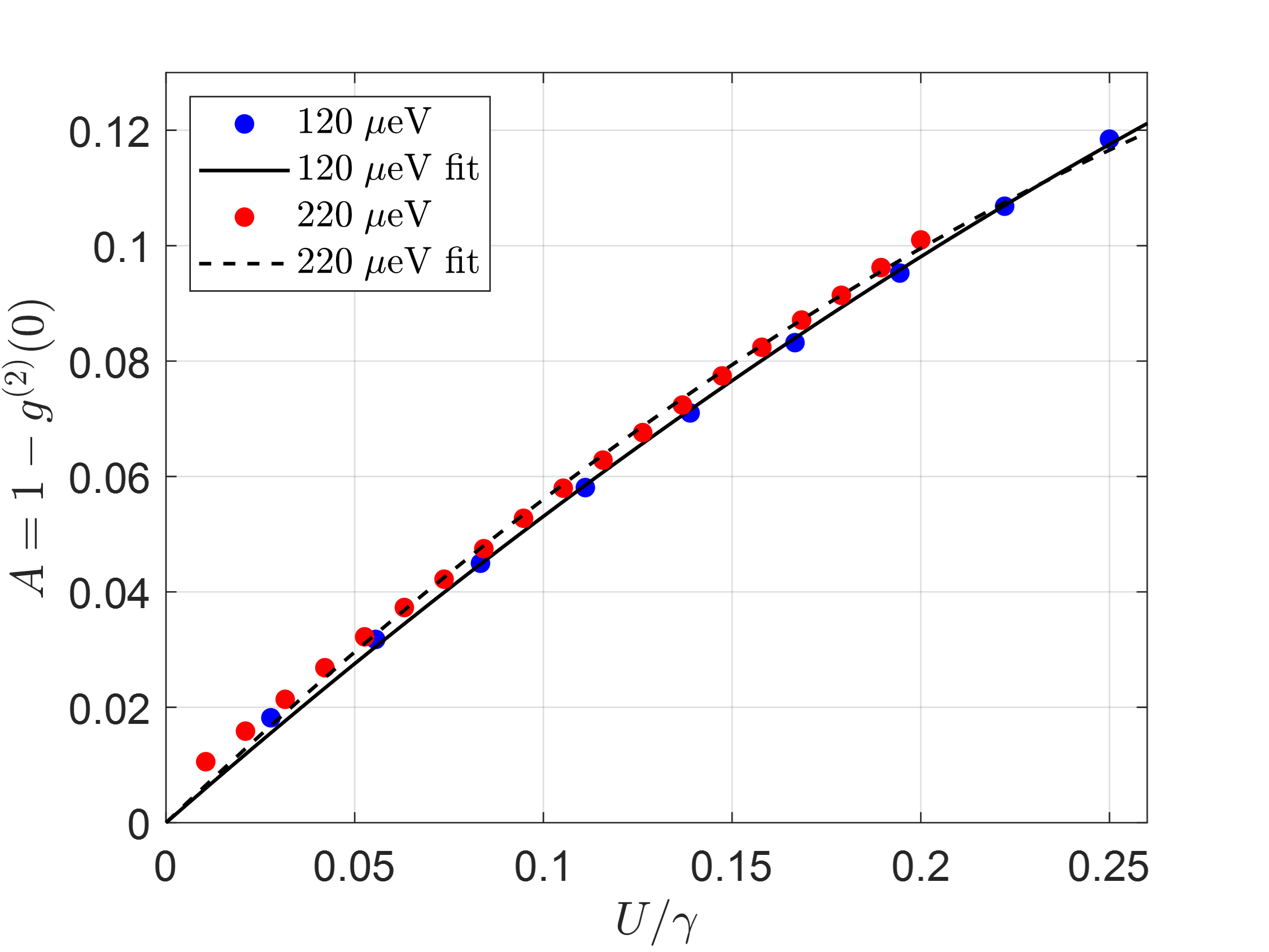}
    \caption{\textbf{Result of the blockade model.} $1-g^{(2)}_{0,min}$ as a function of $U_{dd}/\gamma$, for loss values of $\gamma=120$, $220$~$\mu$eV along with fits to Eq. \ref{eq:fit}.}
    \label{fig:supp_kappa}
\end{figure}

This model was calculated for two values of $\gamma=120,220$~$\mu$eV and a sweep over interaction strengths to obtain a value for $\kappa$. Fig. \ref{fig:supp_kappa} shows the resulting value of $1-g^{(2)}_{0,min}$ for various interactions for the two loss values along with a fit to: 

\begin{equation}
   1-g^{(2)}_{0,min} = \kappa \frac{U_{dd}}{\gamma}+b \left(\frac{U_{dd}}{\gamma}\right)^{2}.
   \label{eq:fit}
\end{equation}

The fit is robust and independent of $\gamma$ for the value ranges of our interest and yields values of $\kappa=0.61\pm 0.01$,  $b=-0.56\pm0.05$ with agreement with previous results \cite{delteil2019towards}. This expression, along with the value of $\gamma$, can now be used to extract $U_{dd}$. The density can then be calculated as: $n=\frac{2}{\tau_{p}v_{g}w}$ and be used to find the interaction strength $g=U/n$. The independence of $\kappa$ on $\gamma$ suggest that our calculation of the interaction strength is robust against inaccuracies in extraction of $\gamma$, because for a given dip, the extracted $U_{dd}$ would change linearly with $\gamma$ but will also cause an inverse change in $n$ (because of the minimal possible pulse width) that cancel each other, making the blockade condition weakly dependent on $\gamma$ as shown in Eq. \ref{eq:constraint}. Moreover, the negative value of $b$ means that our first order approximation will result in an under-estimation of the interaction strengths by roughly $10\%$. The pulse width, $\tau_{p}$ used in our calculation is taken solely from the Fourier-limit of the linewidth, but in principle should be bigger due to the finite spatial size of the laser spot, $\delta$, which, after convolution, will give an effective temporal width of $\tilde{\tau}_{p} = \sqrt{\tau_{p}^{2}+\delta/v_{g}}$ where $v_{g}$ is the group velocity and Gaussian functions were assumed for both the spatial and temporal laser profiles (which will further decrease the density).

\subsection*{Supplementary Text}

\subsubsection*{Blockade radius and WG width analysis}

The interaction strength, $g_{dd}$ and the linewidth, $\gamma$ are the fundamental parameters for determining, the blockade radius $R_b$ and the constraints on its dependence dipole length, WG width and excitonic fraction in our particular system. The blockade radius, $R_{b}$, is simply the maximal distance between two polariton excitations at which the repulsive interaction still result in a blueshift comparable to the linewidth:
\begin{equation}
    U(R_{b}) = \gamma.
\end{equation}
Using the interaction-density relation, $U=g_{dd}n$, where $n=2/A$ for a mode area $A$, and the connection $A=\pi R_{b}^{2}$ we get:
\begin{align}
    \gamma &= g_{dd} \frac{2}{\pi R_{b}^2} \label{eq:basic}  \\
    R_{b} &= \sqrt{\frac{2g_{dd}}{\pi \gamma}}
\end{align}
Which, when calculated for our system, yields $R_{b}= 4-5$~$\mu$m. \\
Next we explicitly derive the constraint on our rectangular WG having a width, $w$,  a dipole length, $d$, and the excitonic fraction, $|\chi_{X}|^{2}$ presented in Eq. \ref{eq:constraint}. Coming back to Eq. \ref{eq:basic} but now considering the rectangular mode area of our system that depends on the WG width, the pulse temporal width, $\tau_{p}=\hbar/\gamma$ and the group velocity $v_{g}$:
\begin{equation}
    \frac{\gamma}{g_{dd}} = \frac{2}{v_{g}\tau_{p}w} = \frac{2}{v_{g}w \hbar/\gamma}
\end{equation}
which can be written as $w = 2g_{dd}/\hbar v_{g}$. Plugging in the linear dependence of the interaction strength on the dipole \cite{liran2024electrically} and the quadratic dependence on the excitonic fraction yields:
\begin{equation}
    w = \frac{2g_{dd}^{ex}}{v_{g}\hbar} \frac{d}{d^{ex}}\frac{|\chi_{X}|^{4}}{|\chi_{X}^{ex}|^{4}}
\end{equation}
where $g_{dd}^{ex},d^{ex},\chi_{X}^{ex}$ are specific corresponding values such as those used in our experiment. We can further replace the general group velocity with $v_{g}=v_{p} (1-|\chi_{X}|^{2}$) with $v_{p}$ the photon velocity (see previous section). The expression can then be rewritten as:
\begin{equation}
    \frac{w \left(1-|\chi_{X}|^{2} \right)}{d|\chi_{X}|^{4}} = \frac{2g_{dd}^{ex} \left( 1-|\chi_{X}^{ex}|^{2} \right)}{\hbar v_{g}^{ex} |\chi_{X}^{ex}|^{4} d^{ex}}
\end{equation}
which gives a unit-less relation between the dipole length and the WG mode width along with its dependence on the excitonic fraction as a function of a constant that from our experimental data is $\sim50$.

\clearpage %

\end{document}